# Statistical Engineering: An Idea Who's Time Has Come?

Roger W. Hoerl        Ronald D. Snee
Union College         Snee Associates, LLC

October 23rd, 2015

**Abstract**

Several authors, including the American Statistical Association (ASA), have noted the challenges facing statisticians when attacking large, complex and unstructured problems, as opposed to well-defined textbook problems. Clearly, the standard paradigm of selecting the one "correct" statistical method for such problems is not sufficient; a new paradigm is needed. Statistical engineering has been proposed as a discipline that can provide a viable paradigm to attack such problems, used in conjunction with sound statistical science. Of course, in order to develop as a true discipline, statistical engineering needs a well-developed theory, not just a formal definition and successful case studies. This article documents and disseminates the current state of the underlying theory of statistical engineering. Our purpose is to provide a vehicle for applied statisticians to further enhance the practice of statistics, and for academics interested in continuing to develop the underlying theory of statistical engineering.

## 1. INTRODUCTION: THE CHALLENGE OF LARGE, COMPLEX AND UNSTRUCTURED PROBLEMS

Many of the world's most challenging statistical problems are large, complex and unstructured. Applications in such areas as genomics, public policy, and national security often present significant challenges, even in terms of precisely defining the specific problem to be solved. For example, in obtaining and utilizing data to protect national security, one could perhaps develop an excellent system in terms of surveillance and threat identification, but one that would result in essentially a police state with no privacy or individual rights. Few would consider this a successful or desirable system. Similarly, the system currently in place to approve new pharmaceuticals in the US involves a series of clinical trials and analyses, guided by significant subject matter knowledge, such as in identifying likely interactions. No single experimental design or statistical analysis results in a new approved pharmaceutical. Further, the system must balance the need for public safety with the urgent need for new medications to combat deadly diseases such as Ebola. The problem is complex!

Conversely, most problems in statistics textbooks are narrow, well defined, and simply require application of one "correct" statistical method, typically the one currently being studied. The need for a broader approach to teaching statistics students to solve complex problems has been recognized by others, such as Meng (2009), who discussed a recently developed course at Harvard, Stat 399. This course "...emphasizes deep, broad, and creative statistical thinking instead of technical problems that correspond to a recognizable textbook chapter." Tukey (1962, p.13) identified this issue decades ago, noting: "Far better an approximate answer to the *right* question, which is often vague, than an *exact* answer to the wrong question, which can always be made precise."

We argue that most real "Big Data" problems are large, complex and unstructured, in that they often involve multiple data sets collected under different conditions, and in fact, the fundamental problem to be solved is frequently not clear (Hoerl et al. 2014). For example, for data competitions posted on kaggle.com is the real problem to find an optimal model to predict a holdout data set, or to develop subject matter knowledge about the phenomenon of interest, or to guide intervention in the system to achieve enhanced results in the future? Such complex problems can rarely be solved with one method, and usually require a sequential approach to link and integrate multiple tools, guided by existing subject matter knowledge.

Along these lines, the ASA recently published a policy statement on data science that states: "New problem-solving strategies are needed to develop 'soup to nuts' pipelines that start with managing raw data and end with user-friendly efficient implementations of principled statistical methods and the communication of substantive results." (https://www.amstat.org/misc/DataScienceStatement.pdf) Clearly, ASA recognizes the need for something broader than a new statistical technique or method, i.e., for sequential approaches involving multiple methods and technologies, to address modern data science problems.

## 2. STATISTICAL ENGINEERING PROVIDE A VIABLE PARADIGM

Statistical engineering has been proposed as providing an alternative paradigm to guide integration of multiple statistical methods to address large, complex and unstructured problems (Hoerl and Snee 2010a, 2010b, 2010c, Snee and Hoerl 2010). Statistical engineering was initially defined in this journal as: "The study of how to best utilize statistical concepts, methods, and tools, and integrate them with information technology and other relevant disciplines, to achieve enhanced results." (Hoerl and Snee 2010a, p.12). We should note that Eisenhart (1950) first published the term statistical engineering, although Eisenhart's focus was more on what we would today call engineering statistics – the application of statistics to engineering problems.

While elaborations of statistical engineering have appeared in the literature, and numerous applications have been published (Anderson-Cook and Lu 2012), an underlying theory of statistical engineering has proven elusive. Snee and Hoerl (2010) is virtually the only published paper to formally discuss the underlying theory. This dearth is partially due to the nature of statistical engineering, which is not inherently mathematical, and partially due to the fact that formal study of statistical engineering as a discipline in its own right is relatively new.

Certainly, the study of how to link and integrate multiple statistical tools to achieve enhanced results goes back to the origins of statistics. However, we feel that most of these applications have been "one-offs", with little theory documented on how to attack such problems in general. We argue that the literature of our discipline has focused much more heavily on individual tools than on strategies for linking and integrating multiple tools to solve complex problems.

**2.1 Why Statistical Engineering is Needed**

In his landmark book on the historical development of science, Kuhn (1962) pointed out that the need for a new way of thinking, i.e., a new paradigm, occurs when the number of problems not solved by the current paradigm is so large that a new approach is clearly required. We believe that this is true today for the statistics profession. In our view, a different paradigm is needed to deal with large, complex and unstructured problems, versus narrow technical problems that can be solved with one technique. This is especially true for Big Data problems that often require integration of multiple data sets collected under different circumstances (Hoerl et al. 2014). While statistical engineering itself is not new, the documentation, clarification, and elaboration of statistical engineering - especially it's underlying theory - is both new and required, in our opinion.

Per Kuhn, there is a growing consensus within the statistics profession that this issue of complex problems needs further attention. For example, the recently published ASA guidelines for undergraduate statistical science programs (Chance et al. 2015, p.6) note:

> Undergraduates need practice using all steps of the scientific method to tackle real research questions. All too often, undergraduate statistics majors are handed a "canned" data set and told to analyze it using the methods currently being studied. This approach may leave them unable to solve more complex problems out of context,

> especially those involving large, unstructured data.... Students need practice developing a unified approach to statistical analysis and integrating multiple methods in an iterative manner.

Of course, it may not be obvious to readers of the ASA report just how to go about: "...developing a unified approach to statistical analysis and integrating multiple methods in an iterative manner". After all, these concepts are rarely explicitly taught in academic courses or emphasized in textbooks, as noted by Notz (2012). Anderson-Cook et al. (2005), which describes a course in problem solving introduced at Virginia Tech, is a positive counter-example. Despite a few positive counter-examples, however, a new paradigm is clearly needed to fill in the gap.

It should not go unnoticed that such big problems provide significant opportunities to statisticians. In addition to opportunities to demonstrate tangible impact on projects with high visibility, such problems also provide the real leadership opportunities that statisticians often seek, but frequently have difficultly finding (Rodriguez 2012). No profession has a monopoly on large, complex problems, and many statisticians have developed skills in problem solving, making them logical candidates for project leadership on such problems.

We recognize that good statisticians have utilized what we are calling statistical engineering to solve complex problems since the inception of statistics. For example, as far back as the middle of the 20th century Box and Wilson (1951) proposed a sequential approach to empirically optimize response surfaces involving multiple rounds of experimental designs and regression analyses. They did not propose an optimal experimental design, or a preferred model per se, but rather a novel approach to linking and integrating multiple tools to solve a real problem.

As groundbreaking as this paper was, however, it did not provide general principles or guidelines to help other researchers solve large, complex problems that might be very different from optimizing response surfaces. In other words, it did not elucidate the paradigm of statistical engineering, or its underlying theory. For statistical engineering to develop as a true discipline, however, its underlying theory must be more formally documented and then developed over time.

**2.2 What is Statistical Engineering?**

The definition of statistical engineering, previously noted, is: "The study of how to best utilize statistical concepts, methods, and tools, and integrate them with information technology and other relevant disciplines, to achieve enhanced results." Some key words and phrases in this definition warrant elaboration. First of all, statistical engineering is defined as "the study" of something, i.e., a discipline, not a tool or technique. As with any other engineering discipline, it utilizes existing concepts, methods, and tools in novel ways to achieve novel results, typically through integration of multiple tools.

Various definitions of engineering are available (e.g., [http://www.merriam-webster.com/dictionary/engineering](http://www.merriam-webster.com/dictionary/engineering)). These accepted definitions generally emphasize utilization of existing science and mathematics in novel ways to achieve novel results. An old saying in the engineering community is: "An engineer is someone who can accomplish with $1 what any fool can accomplish with $2." Conversely, most definitions of science involve the study and advancement of the fundamental knowledge of the physical word (e.g., http://www.merriam-webster.com/dictionary/science). While science emphasizes development of new fundamental knowledge, engineering finds novel ways to utilize this knowledge for the benefit of humankind.

Former MIT President Susan Hockfield (2010) made a similar point, noting that around the dawn of the 20th century, physicists discovered the basic building blocks of the universe (i.e., the periodic table), which could be considered a "parts list". However, it was engineers who figured out how this parts list could be put to best use, subsequently driving the electronics and computer revolutions. Similarly, Hockfield noted that biologists had recently discovered the basic building blocks of life (the human genome), another "parts list", but now engineers are subsequently finding creative ways to utilize this parts list, such as in personalized medicine.

We argue that statisticians have developed an excellent toolkit over the past centuries, which could be considered another "parts list", but that insufficient thought has gone into the engineering problem of how best to integrate multiple tools in creative ways to solve complex problems. At least, insufficient thought has gone into documenting the underlying theory of how to approach this engineering problem in general.

Integration is a key word in the definition of statistical engineering, including integration with other disciplines, especially information technology. In fact, one might argue that the emerging discipline of data science is essentially the integration of statistics with information technology. The ASA policy statement on data science noted previously makes essentially this same point. We agree that information technology is a key discipline with which integration is frequently required for large, complex and unstructured problems. Conversely, narrow textbook problems can typically be solved via the application of one "correct" tool; integration is rarely required.

The phrase "achieve enhanced results" is key in the sense that statistical engineering is inherently tool agnostic. That is, it neither promotes frequentist nor Bayesian methods, neither classical nor computer-aided experimental designs, neither parametric nor non-parametric (or quasi-parametric) approaches, and so on. Rather, as an engineering discipline, its "loyalty" is to solving the problem – generating results, rather than to pre-determined methods. Tools are of course important, but within a statistical engineering paradigm they would be chosen based on the unique nature of the problem, to generate results. Various philosophies and tools sets may be employed, such as integration of frequentist and Bayesian methods, for example.

**2.3 What is Theory?**

Within the statistical literature, the word "theory" typically refers to mathematical statistics, i.e., theorem-proof, or mathematical derivations of distributions, for example. However, it is well known that theory has a much broader meaning, not only outside of statistics, but within statistics as well. For example, Madigan and Stuetzle, in their discussion of Lindsay et al. (2004, p.409), made essentially this same point: "The issues we raise above have nothing to do with the old distinction between applied statistics and theoretical statistics. The traditional viewpoint equates statistical theory with mathematics and thence with intellectual depth and rigor, but this misrepresents the notion of theory. We agree with the viewpoint that David Cox expressed at the 2002 NSF Workshop on the Future of Statistic that 'theory is primarily conceptual,' rather than mathematical."

As with engineering and science, many definitions of the word theory are possible (e.g., http://www.merriam-webster.com/dictionary/theory). However, reasonable and accepted definitions typically state something similar to: "a coherent group of general propositions used to explain a phenomenon". Obviously, there is no explicit requirement in such definitions for mathematics to be involved, although it often is.

The underlying theory of physics, for example, involves considerable mathematics, but of course not all of the theory of physics is mathematical. If it were, physics would be considered a sub-field of mathematics. In our opinion, one of the reasons so many professionals still consider statistics a sub-field of mathematics is because statistics as a profession has not adequately articulated that part of statistical theory that is not mathematical in nature.

Deming (1986, 2000) also spoke quite a bit about theory. For example, he is credited with the following comments about the role of theory in data-based learning (Snee 2012):

- Without theory there is nothing to revise…..no learning.
- Information is not knowledge….Knowledge comes from theory.
- Experience teaches nothing unless studied with the aid of theory.
- An example teaches nothing unless studied with the aid of theory.
- A theory may be complex. It maybe simple. It may be only a hunch, and the hunch may be wrong. We learn by acceptance or by modification of our theory, or even abandoning it and starting over.

Our point in reference to Deming, as well as Madigan and Stuetzle, and also Cox, is that the importance of theory, particularly conceptual as opposed to mathematical theory, has been well known within the statistics profession for decades. Further, theory is needed in order to learn from experience or case studies. We therefore argue that for statistical engineering to serve the statistics profession as a legitimate discipline, a solid theoretical foundation is needed.

## 3. THE UNDERLYING THEORY OF STATISTICAL ENGINEERING

For an initial depiction of the underlying theory of statistical engineering, we consider how it fits within the broader view of statistics as a system, the key principles of the theory as they relate to large, complex and unstructured problems, and also a high-level model to guide application of these principles to real problems, based on previous research as well as our own experiences.

**3.1 The Big Picture View**

In order to depict the theory of statistical engineering, we must consider how it relates to more familiar and well-documented aspects of statistics, such as statistical science, statistical practice, statistical thinking, and the individual tools and methods. That is, we need a conceptual model of how these aspects integrate to form the broader discipline of statistics as we know it today.

It is hard to find a generally accepted definition of the term "statistical science", so referring back to Hockfield's (2010) discussion of science and engineering, we will use the term statistical science to refer to the "parts list", that is, the study of the fundamental laws of statistics and properties of individual methods, in contrast to statistical engineering, which focuses more on integration of multiple methods, and discovering how these parts might be more effectively utilized to solve large, complex problems. Obviously, statistical science and statistical engineering are both required, and should interact synergistically, as do chemistry and chemical engineering.

Figure 1 (Snee and Hoerl 2010a) depicts the statistics discipline as a system, with strategic, tactical, and operational levels, each of which has both a theoretical and applied element. Per Meng (2009), the strategic aspect is statistical thinking; how we think about statistics itself, and its relationship with other disciplines. The operational aspect is where the "rubber hits the road", i.e., the actual methodologies of statistics, such as modeling techniques, experimental design, and so on.

However, we have found a gap between the higher-level principles of statistical thinking and utilization of the individual tools. That is, how should researchers or practitioners utilize statistical methods in such a way as to be consistent with the principles of statistical thinking? In our experience, we have found that too often they are not. For example, the recent ASA guidelines on undergraduate statistical education, quoted previously, noted: "Students need practice developing a unified approach to statistical analysis and integrating multiple methods in an iterative manner."

We propose that statistical engineering can fill this gap and serve as the tactical aspect of the discipline, linking the individual methods with the fundamental principles of statistical thinking, such as the need for a unified approach. Once the strategic principles are understood, statistical engineering can provide frameworks based on these principles to help link and integrate individual methods.

We believe that this applies to both research and applications. Each of the three aspects – statistical thinking, statistical engineering, and the methods and tools – needs to have a well-defined theory, as well as experienced-based principles guiding application. As a trivial example, most statistical inference methods assume a random sample, while experienced practitioners know that random sampling is the exception rather than the norm in most real applications.

**Figure 1**

**Statistics as a System**

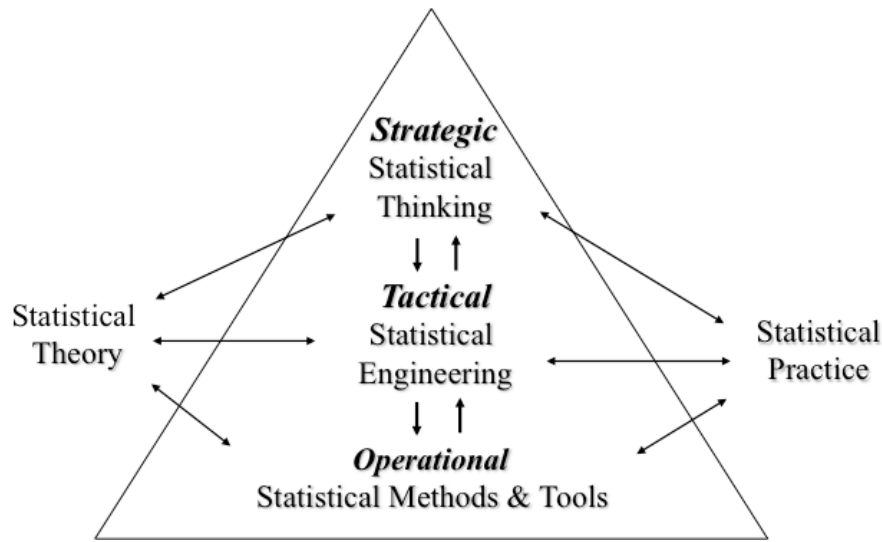

**3.2 Core Principles of Statistical Engineering**

In our view, statistical engineering is not typically needed to address straightforward problems, or even complex problems that are narrow and well defined. Rather, it is particularly needed for large, complex and unstructured problems. Therefore, the following discussion of core principles underlying statistical engineering assumes that the problem in question is large, complex and unstructured. We document some critical elements of what the profession has learned over the past centuries, and on which it is in general agreement, as to the key factors that determine the ultimate success or failure of statistical problem solving. This is in some sense a "theory"; i.e., "a coherent group of general propositions used to explain a phenomenon".

For example, it has been pointed out (Hoerl et al. 2014) that some of the egregious failures of Big Data analytics, which tend to be much less publicized than the successes, were in large part due to lack of understanding of fundamentals. One common example is the often unstated assumption of data quality, that is, the naive assumption that data are "innocent until proven guilty". Another would be the belief that data can be properly analyzed without any understanding of the problem context, or knowing how the data were collected.

Most experienced statisticians learn these types of pitfalls "on the job", often through making their own mistakes. At this point, they might be considered principles of statistical practice, or applied statistics. However, we argue that such principles can be studied, documented, debated and enhanced over time, and formally taught to students. Under these circumstances, they would be considered a theory. The logical expectation in most disciplines is that theory and practice should gradually converge over time; we believe that the same should be true of statistics.

Note that we make no claims of originality relative to these principles; rather, we are documenting the most important learnings from the statistical literature, conferences, interactions with colleagues, as well as our own experiences. In our view, the most critical principles of statistical engineering applied to large, complex and unstructured problems can be loosely grouped into the five major categories listed in Table 1.

**Table 1**

**Core Principles of Statistical Engineering**

1. Understanding of the problem **context**
2. Development of a problem solving **strategy**
3. Consideration of the data **pedigree**
4. Integration of sound **subject matter theory** (domain knowledge)
5. Utilization of **sequential approaches**

**3.2.1 Context**

Few experienced statisticians in academia, government, or the private sector would argue that understanding the context of the problem being addressed prior to applying statistical methods is unnecessary. Yet, we are seeing a significant increase in the number of data competitions globally, often with minimal opportunity for competitors to properly understand the context of the problem. For example, some kaggle.com competitions do not allow participants to know the actual names of the variables in the data set, in order to protect confidentiality. Rather, participants are only provided a high-level view of the problem being addressed, and the types of variables included in the data set, along with data labeled generically as "x1", "x2", and so on. Clearly, someone thinks an actionable model can be developed without a proper understanding of the problem context, or the data would not have been posted online!

A sound understanding of problem context leads to proper problem definition and scope, and to appropriate statistical goals. The best statistical solution may not be the best technical or business solution. For example, for many real data sets, development of the "best" model is not an appropriate objective. Rather learning as much as possible from the current data, in order to obtain more relevant data, perhaps through a designed experiment, may be a more appropriate initial objective. This could be the case because a judgment sample or convenience sample was used in the initial data collection effort, or because of other data quality issues.

### 3.2.2 Strategy

The second point refers to the value in developing an overall strategy to attack large, complex and unstructured problems. As noted by Meng (2009), textbook problems typically require identification of the correct statistical method, and subsequently to the proper application of this method to the data provided. However, with complex problems there is no one correct method, and in fact, more than one method is likely to be required. No appropriate data may be initially available. Therefore, some serious thought is needed in terms of how to approach the problem in the first place, including identification of the most appropriate data available. A well-defined strategy is typically much more effective than a "flurry of activity".

As a positive example, we argue that Box and Wilson (1951), mentioned previously, provided an overall strategy for attacking the problem of process optimization over half a century ago. Unfortunately, we do not feel that there is adequate discussion of strategies for problem solving in statistics textbooks or courses, in academia, government or the private sector. The more complex and unstructured the problem is, the more likely a multi-step approach will be required to adequately address it. Of course, the strategy needs to be based on the specific problem on hand, and the specific objectives. No single strategy is appropriate for all problems (Hoerl and Snee 2013). In the next section we discuss one approach to developing such an overall problem solving strategy, based on the specific problem at hand.

**3.2.3 Data Pedigree**

The third key point notes the importance of understanding the "pedigree" of any data analyzed. Just as the pedigree of a show dog or racehorse is critical to determination of the animal's value, so an understanding of where the data came from, and specifically how they were collected, are critical to determining the value of the data. As noted, it is disturbing to us that many data competitions provide minimal consideration of the pedigree of data provided. In our collective experience analyzing real data sets, we have found that many data sets are simply not worth spending a lot of time analyzing.

We are not only referring to blunders in the data set or missing values, but to such things are biased sampling, inappropriate timeframes for data collection, inappropriate selection of variables, and poor measurement processes. We acknowledge that our observation is not new. Quoting again from Tukey (1986 pp.74-75): "The combination of some data and an aching desire for an answer does not ensure that a reasonable answer can be extracted from a given body of data."

Virtually no real data set, for example, could be considered a true random sample from the population of interest. In the case of financial applications, such as default prediction or credit scoring, the population of interest usually consists of future observations. The conclusion, often accurate, that current data are not appropriate to answer the fundamental question at hand does not seem to be an acceptable response in virtually any data competitions of which we are aware.

### 3.2.4 Subject Matter Knowledge

A positive trend in the statistics profession in the past couple of decades has been the growth of interdisciplinary research. For example, one of the motivations for launching the National Institute for Statistical Science (NISS) in 1990 was to facilitate interdisciplinary research (see www.niss.org). Two obvious reasons for this trend are the facts that subject matter experts typically "own" the original problems of interest to society, and that proper application and interpretation of statistical methods require significant subject matter expertise. This second reason is particularly true when the goal is to go beyond "yes/no" statistical significance decisions to the question of practical significance, and to reconsideration of the underlying theory of the phenomenon of interest, based on statistical analyses.

While the value of subject matter theory is well understood by much of the statistical community, we continue to see too many examples where it is clearly not well understood, within the broader community that utilizes statistical methods, including the fields of "data analytics" and Big Data. For example, there have certainly been numerous evaluations and opinions as to the root causes of the 2008/2009 financial crisis, and in particular, why the financial risk models in place at the time by and large failed to give adequate warning (e.g., Thomas et al. 2011). Some went so far as to question the Nobel Prize awarded to Merton, Black, and Scholes for their fundamental developments in risk modeling (Haug and Taleb 2011).

In our view, however, very little has been said about the fact that many – perhaps most – of the individuals developing, utilizing, and maintaining financial risk models at that time were mathematicians, statisticians, physicists, engineers, and so on, with strong backgrounds in mathematics and perhaps modeling, but with limited backgrounds in financial theory (Biello 2011).

As is obvious to many, sound subject matter knowledge is needed not only to interpret data analyses appropriately, but also to determine what data should be obtained in the first place, to evaluate the data pedigree, and especially to consider if the existing subject matter theory needs to be modified based on the current analysis. Unfortunately, we feel that students could read many popular statistics textbooks without ever learning this critical point. Of course, statisticians, or those trained in statistical methods, serve a critical role in proper use and understanding of the methods, providing further evidence of the value of interdisciplinary statistical projects.

**3.2.5 Sequential Approaches**

The last point in Table 1 refers to the fact that the majority of successful statistical projects are sequential in nature, involving a series of data collection and analysis steps, each based on what was learned on previous steps. Again, we claim no originality to this point; Box et al. (1978) emphasized this principle decades ago. Approval by the FDA of new pharmaceuticals via a series of clinical trials is a classic example of this model in practice. Going back to Meng (2009), however, the vast majority of examples provided in statistical textbooks, and we would argue, statistical publications, involve one technique applied to one data set, i.e., they are "one-shot studies", giving the impression that this is how statistics is supposed to be researched and applied.

A main advantage of sequential studies is the principle that "hindsight is 20/20"; that is, after the fact we know exactly what we should have done initially. Box (1993) quoted R.A. Fisher as stating: "The best time to design an experiment is after you have done it." With a sequential approach, it becomes possible to use hindsight to our advantage, guiding the next round of experimentation or analysis. Such an approach leads naturally to linking and integrating multiple methods, a key aspect of attacking large, complex and unstructured problems. We further argue that a sequential approach, involving repeated application of both deductive (theory to observation) and inductive (observation to theory) thinking, enables a balanced emphasis on creatively developing new theories as well as testing existing theories. In our view, this is when statistics becomes most potent as a discipline.

**3.3 A High-level Model for Applying Statistical Engineering**

We have found that statistical engineering applications – the linking and integration of multiple tools to solve large, complex and unstructured problems – tend to follow a somewhat repeatable process. Further, in order to teach students how to address such problems, rather expecting them to learn by "trial and error", some methodology or approach must be proposed.

DiBenedetto (2014) researched problem-solving approaches from a wide spectrum of disciplines, and found that the approaches from these diverse disciplines shared much in common. Figure 2, taken from DiBenedetto et al. (2014), provides a high-level model for applying statistical engineering to attack large, complex and unstructured problems based on this research. We should note that this is not intended as a "cookbook", or "7 easy steps to problem solving", but rather as a framework that must be tailored to the unique context of each individual problem to develop a strategy for addressing that problem. As a model, of course, it can and should be improved upon over time.

**Figure 2**

**Typical Phases of Statistical Engineering Projects**

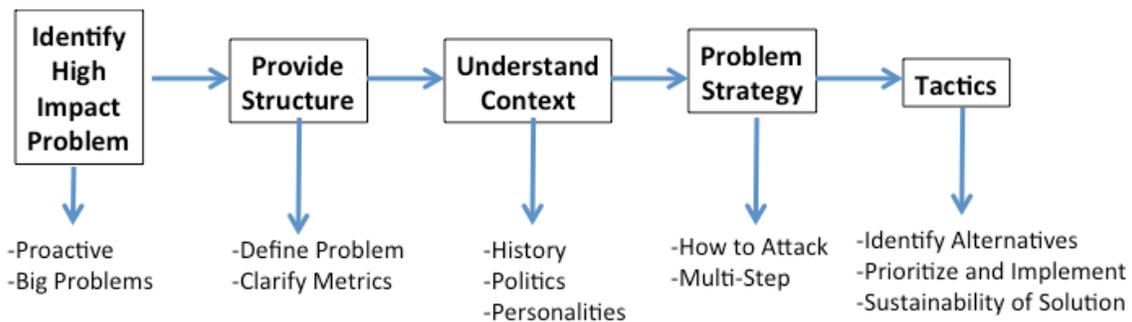

The first phase involves identification of high-impact problems. For many practicing statisticians, as well as those in academia, the most impactful problems rarely come knocking on the door, but require proactive effort to find. As noted by Lin (2014), "Finding a good problem is harder than finding a good solution." Further, non-statisticians may not see the statistical aspect of the problem, especially if no data have yet been collected. That is, many professionals may wait until a narrow technical problem has been defined to add a statistician to the team.

Since by definition large, complex and unstructured problems are unstructured, the next step is to provide enough structure to move forward. For example, the first author led a team attempting to develop a corporate default predictor for GE Capital (Neagu and Hoerl 2005). However, upon discussion of the problem with subject matter experts, it became clear that there is no commonly accepted definition of the word "default" in financial circles. Clearly, predicting something that is not defined is complex! In addition to defining default, other required structure included development of metrics to quantify success or failure in the project, such as defining what would or would not constitute adequate timing of prediction, appropriate measures of Type I and Type II errors, and so on.

Once the problem is reasonably structured, a logical next step is to delve into the context of the problem; its history, politics and personalities involved, why it has not been resolved previously, and so on. For narrow technical problems, such context is not critical, but for complex problems it typically is. That is, there are reasons why the problem has not been solved previously. For example, much research has gone into default prediction, and there are good reasons why it remains today essentially an unsolved problem. To give an example from a completely different discipline, complex health problems, such as the global HIV/AIDS pandemic, cannot be resolved if one only looks at the epidemiological aspects on HIV, and is not keenly aware of the local social, political, cultural, and religious issues surrounding the pandemic (Hoerl and Neidermeyer 2009).

Once an important problem has been identified, adequate structure developed, and the context understood, one is finally in a position to develop an overall strategy for attacking the problem. For complex problems, this will almost always involve multiple steps and multiple methods, many of which may be non-statistical in nature, such as those from information technology. Typically, the initial data, or perhaps better data than is currently available, will need to be obtained. Often, disparate data sets will need to be integrated, especially in Big Data problems. The strategy may involve breaking the overall problem into sub-problems that can be more easily addressed, as was the case in the default prediction project.

Within the overall strategy there will be individual tactics selected and applied, such as identifying alternative approaches to a given sub-problem, and prioritizing these alternatives. With complex problems, simply listing all viable potential approaches can be an important step towards solution, and is commonly applied in many disciplines (DiBenedetto 2014). Similarly, consideration of how a final solution might be maintained is a tactic that should not be overlooked. Many complex problems are "solved", only to have the solution gradually fade away, and the original state return. Weight loss programs and smoking-cessation programs would be obvious examples from everyday life. In the default prediction project a "control plan" utilizing an ongoing measure of model accuracy that incorporated censored data methods from reliability was used to evaluate the need to retune the model over time (Neagu and Hoerl 2005).

**3.4 Enhancing Statistical Engineering Theory**

The theory of statistical engineering will be refined and grow as the approach is used in practice, and is the subject of more academic research. To date, statistical engineering has been defined, its principles developed at a basic level, and its role relative to statistical thinking and also statistical methods and tools has been identified (Figure 1). Several case studies utilizing statistical engineering have been published (Anderson-Cook and Lu, 2012). The body of knowledge is growing. Ongoing assessment of the theory as guidance for application of statistical engineering to large, complex and unstructured problems will strengthen and add to the theory over time.

## 4. SUMMARY AND PATH FORWARD

Several authors, including those of a recent ASA policy statement, have noted the need for developing statistical approaches to large, complex and unstructured problems, that is to problems that do not "correspond to a recognizable textbook chapter." Clearly, individual statistical methods, no matter how powerful, will not suffice. Rather, a new paradigm is needed to link and integrate multiple methods in an overall strategy to solve such problems. Fortunately, the history of the statistics profession provides many examples of doing just that. Unfortunately, however, underlying theory as to how the individuals involved approached the problem and developed the solution is, in our opinion, lacking.

Further, we agree with the ASA guidelines for undergraduate education that: "Students need practice developing a unified approach to statistical analysis and integrating multiple methods in an iterative manner." We feel that a formal discipline of statistical engineering can provide a paradigm to help address both of these issues. To date, numerous examples and case studies have been published, such as Hare (2011), Anderson-Cook and Lu (2012), which is a special edition on statistical engineering including several case studies, Steiner and MacKay (2014), and Snee et al. (2016). More published case studies are clearly needed.

In order for statistical engineering to emerge as a formal discipline, however, an underlying theory must also be documented and then revised, developed and improved upon over time. This article provides a start to such documentation, admittedly a coarse and unrefined one. Only by publishing an initial documentation of the theory can it be subsequently refined and improved over time.